\documentclass[a4paper,10pt]{amsart}

\newcommand{\field}[1]{\mathbb{#1}}
\newcommand{\Lie}{\mathcal{L}}
\newcommand{\R}{\field{R}}
\newcommand{\N}{\field{N}}
\newcommand{\Z}{\field{Z}}

\DeclareMathOperator{\inter}{int}
\DeclareMathOperator{\rank}{rank}
\DeclareMathOperator{\diag}{diag}

\newtheorem{theorem}{Theorem}
\newtheorem{lemma}{Lemma}

\begin{document}

\title{On the Theory of Killing Orbits in Space-Time}

\author{G.S. Hall}

\address{Department of Mathematical Sciences\\
University of Aberdeen, Aberdeen AB24 3UE, Scotland, UK.}

\date{}

\begin{abstract}
This paper gives a theoretical discussion of the orbits and
isotropies which arise in a space-time which admits a Lie algebra
of Killing vector fields. The submanifold structure of the orbits
is explored together with their induced Killing vector structure.
A general decomposition of a space-time in terms of the nature and
dimension of its orbits is given and the concept of stability and
instability for orbits introduced. A general relation is shown
linking the dimensions of the Killing algebra, the orbits and the
isotropies. The well-behaved nature of "stable" orbits and the
possible miss-behaviour of the "unstable" ones is pointed out and,
in particular, the fact that independent Killing vector fields in
space-time may not induce independent such vector fields on
unstable orbits. Several examples are presented to exhibit these
features. Finally, an appendix is given which revisits and
attempts to clarify the well-known theorem of Fubini on the
dimension of Killing orbits.
\end{abstract}

\maketitle

\section{Introduction}

The purpose of this paper is to develop in a reasonably modern way
the theory of Killing orbits in space-times. It is not claimed
that all the results given are new but, at least in the form
presented here, some are new to the present author. It is hoped
that the proofs given are a little tidier and that the conditions
required for the results stated are clearer.

The theory of Killing symmetry is often used in the construction
of exact solutions of Einstein's field equations and is important
in the physical applications of general relativity. However,
certain mathematical problems arise in the study of such
symmetries and it is of some importance for these to be at least
realised and, if possible, rectified. For example, exact solutions
are often found by imposing a certain collection of Killing vector
fields on a space-time with orbits of a certain prescribed
\emph{nature} (timelike, spacelike or null) and \emph{dimension}.
The resulting metrics, not surprisingly, have precisely these
symmetries and orbits. However, they may be extendible to larger
manifolds where the orbit nature and dimension could change. A
less restrictive way of approaching such a problem is to find
general results which link the possible orbit types with the
dimension of the Killing algebra. This will be discussed in more
detail later in the paper. Another potential problem arises from
what the author perceives as the occasional unjustified
(unconscious) assumption that \emph{independent} Killing vector
fields in space-time naturally project to \emph{independent}
Killing vector fields in the induced geometry of any (non-null)
orbit and thus to physical interpretations in this orbit geometry.
That the Killing nature is preserved is clear, but the
independence need not be. Examples will be given later where this
independence assumption fails. It will also be shown that this
assumption is correct provided a type of "stability" condition is
imposed on the orbit and that this condition naturally holds for
the orbits through "almost every" point of the space-time. This
stability property of orbits has some consequences for the nature
of the Killing isotropy at a space-time point on the orbit. Such
isotropy is important for the physical interpretation of the
solution and may depend on the stability of the orbit. This will
also be clarified later. It would be interesting to find a general
physical interpretation of metrics which admit unstable orbits.
There are also some mathematical problems associated with the
orbits arising naturally from the imposed algebra of Killing
vector fields and these will also be reviewed. It is, after all,
important from a physical viewpoint, that such orbits are
sufficiently well behaved for calculus to be used on them. A
further problem occurs with the well-known application of Fubini's
theorem to the possible dimension of the Killing algebra. This is
discussed and resolved in the appendix.

Throughout the paper $M$ will denote the usual smooth (connected,
Hausdorff, 4-dimensional) space-time manifold with smooth Lorentz
metric $g$ of signature $(-,+,+,+)$. Thus $M$ is paracompact. A
comma, semi-colon and the symbol $\Lie$ denote the usual partial,
covariant and Lie derivative, respectively, the covariant
derivative being with respect to the Levi-Civita connection
$\Gamma$ on $M$ derived from $g$. The associated curvature, Weyl
and Ricci tensors will be denoted in component form by
$R^a{}_{bcd}$, $C^a{}_{bcd}$ and $R_{ab}(\equiv R^c{}_{bcd})$. If
$m\in M$, the tangent space to $M$ at $m$ is denoted by $T_mM$. A
2-dimensional subspace $U$ of $T_mM$ is referred to as a 2-space
and is called \emph{spacelike} if each non-zero member of $U$ is
spacelike, \emph{null} if it contains exactly one null direction
and \emph{timelike} if it contains exactly two distinct null
directions. Such a classification is mutually exclusive and
exhaustive of 2-spaces at any $m\in M$. It follows that $U$ is
spacelike if and only if it contains no null members and that $U$
is timelike if and only if it contains a timelike member. The
family of 3-dimensional subspace (3-spaces) of $T_mM$ can be
similarly classified as spacelike, null and timelike, the
definitions for spacelike and null 3-spaces being identical to
those for 2-spaces. A 3-space $U$ is called \emph{timelike} if it
contains at least two (and hence infinitely many) distinct null
directions and this is equivalent to it containing a timelike
member. A non-zero skew-symmetric tensor (\emph{bivector}) $F$ at
$m$ has even (matrix) rank. If its rank is two it is called
\emph{simple} and may be written as $F_{ab}=2p_{[a}q_{b]}$ for
covectors $p$ and $q$ at $m$. The 2-space at $m$ spanned by $p^a$
and $q^a$ is called the \emph{blade} of $F$ and $F$ is called
\emph{spacelike} (respectively, \emph{timelike, null}) if this
blade is spacelike (respectively, timelike, null). If $F$ has rank
four it is called \emph{non-simple}.

\section{Submanifolds and Generalised Distributions on $M$}

In this section only, $M$ can be any (smooth) paracompact
manifold. A subset $N$ of $M$ is called a \emph{submanifold} of
$M$ if $N$ itself is a manifold and if the inclusion map
$i:N\rightarrow M$ is a smooth immersion (i.e. the derived linear
map $i_*$ has rank equal to $\dim N$ at each $m\in M$, or
equivalently, $i_*$ is a one-to-one map $T_mN\rightarrow T_mM$ at
each $m\in N$). The natural manifold topology of $N$ need not
equal the subspace topology on $N$ induced from the manifold
topology of $M$ (but always contains it since $i$ is continuous).
If it does, $N$ is called a \emph{regular submanifold} of $M$
\cite{1}. It is remarked here that any topological property given
to a submanifold $N$ refers to its natural manifold topology and
may not necessarily hold with respect to its induced subspace
topology from $M$ unless $N$ is regular. It is also noted that an
open subset of $M$ has a natural structure as an \emph{open}
submanifold of $M$ of the same dimension as $M$ and is regular. It
is clear that if $N'$ is a submanifold of $N$ and $N$ is a
submanifold of $M$ then $N'$ is a submanifold of $M$. Less obvious
is the fact that if $N'$ and $N$ are submanifolds of $M$, with $N$
regular, and if $N'\subseteq N$, then $N'$ is a submanifold of
$N$. A \emph{subset} $N$ of $M$ may be given more than one
structure as a submanifold of $M$ and these structures may have
different dimensions. Unfortunately, it is not convenient simply
to insist in the definition of a submanifold that it is regular
since many `natural submanifolds' evolve in differential geometry
which are not regular. Regular submanifolds have some pleasant
properties which do not necessarily hold for non-regular
submanifolds. For example, if $M_1$ and $M_2$ are any smooth
manifolds and if $f:M_1\rightarrow M_2$ is a smooth map whose
range $f(M_1)$ lies inside a submanifold $N$ of $M_2$, then $f$,
considered as a map $f:M_1\rightarrow N$, is not necessarily
smooth, or even continuous. (In fact, if it is continuous, it is
smooth.) However, if $N$ is a regular submanifold of $M_2$ then
the map $f:M_1\rightarrow N$ is necessarily smooth. It should be
noted, however, that if a subset $N$ of $M$ can be given the
structure of a regular submanifold of $M$ then its dimension is
fixed (by $M$ and $N$) and it admits no other submanifold
structures of this dimension. Thus such a regular submanifold
structure on $N$ is unique. To see this note that if $N_1$ and
$N_2$ denote the subset $N$ with two regular submanifold
structures, the smooth inclusion map $i:N\rightarrow M$ leads to a
smooth bijective map $i:N_1\rightarrow N_2$ with smooth inverse
and so $N_1$ and $N_2$ are diffeomorphic. A standard property of
immersions can then be used \cite{1}  to show that if $N_3$ is any
submanifold structure on $N$ of the same dimension as $N_1$ then
$N_3$ coincides with $N_1$.

There is structure which is intermediate between submanifolds and
regular submanifolds. A subset $N$ of $M$ is called a \emph{leaf}
\cite{2} if it is a connected submanifold of $M$ and if whenever $T$ is
any locally connected topological space and $f:T\rightarrow M$ a
continuous map with $F(T)\subseteq N$, the map $f:T\rightarrow N$
is continuous. It follows that if $M_1$ and $M_2$ are any
paracompact manifolds and $N\subseteq M_2$ is a leaf and if
$f:M_1\rightarrow M_2$ is a smooth map whose range $f(M_1)$ lies
in $N$ then the map $f:M_1\rightarrow N$ is continuous and hence
smooth \cite{2}. It follows by an argument similar to that for regular
submanifolds that if $N$ can be given the structure of a leaf of
$M$ then its dimension is fixed (by $M$ and $N$) and that it
admits no other submanifold structure of this dimension. Thus such
a leaf structure on $N$ is unique.

It is clear that every connected regular submanifold of $M$ is a
leaf of $M$ and, of course, that every leaf of $M$ is a connected
submanifold of $M$. Neither converse is true. In fact the well
known `irrational wrap' on the torus is a leaf which is not
regular \cite{2} and the `figure of eight' (connected) submanifold
in $\R^2$ \cite{1} is not a leaf.

Another useful result is now available. Let $M_1$ and $M_2$ be
(smooth) manifolds with respective submanifolds $N_1$ and $N_2$
and let $f:M_1\rightarrow M_2$ be a smooth map such that
$f(N_1)\subseteq N_2$. Then the associated map $f:N_1\rightarrow
M_2$ is clearly also smooth (since it involves only an elementary
composition with the smooth immersion $i_1:N_1\rightarrow M_1$)
and if the associated map $f:N_1\rightarrow N_2$ is also smooth
then the differential of $f$, $f_*$, maps the tangent space
$T_mN_1$ to $N_1$ at $m$ into the tangent space $T_{f(m)}N_2$ to
$N_2$ at $f(m)$ \cite{1}. A special case of this result will be
useful in what is to follow. This is that if $M$ is a smooth
paracompact manifold and $N$ is a leaf of $M$ and if
$f:M\rightarrow M$ is smooth and $f(N)\subseteq N$, then $f$ is
smooth and $f_*$ maps $T_mN$ into $T_{f(m)}N$ for $m\in N$. If, in
addition, f is a diffeomorphism then $f_*$ is an isomorphism
$T_mN\rightarrow T_{f(m)}N$ at each $m\in N$.

Now let $A$ be a vector space of global smooth vector fields on
$M$. Define a \emph{generalised distribution} $\Delta$ on $M$ as a
map which associates with each $m\in M$ a subspace $\Delta(m)$ of
$T_mM$ given by
\begin{equation}
m\rightarrow \Delta(m) = \{X(m):X\in A\}.
\end{equation}
There is no requirement that $\dim\Delta(m)$ be constant on $M$
and so $\Delta$ is not necessarily a Fr\"obenius type of
distribution.

For any global smooth vector field $X$ on $M$ there is associated
a collection of local smooth diffeomorphisms (local `flows') on
$M$ each of which is obtained in the following way. Let $m'\in M$.
Then there exists an open neighbourhood $U$ of $m'$ and a real
number $\epsilon>0$ such that through any point $m$ of $U$ there
is an integral curve $c_m$ of $X$ defined on
$(-\epsilon,\epsilon)$ such that $c_m(0)=m$. The maps
$\phi_t:U\rightarrow M$ defined by $\phi_t(m)=c_m(t)~(t\in
(-\epsilon,\epsilon))$ are the required local smooth
diffeomorphisms \cite{3,4}. The set of all such local
diffeomorphisms obtained in this way for each $m'\in M$ and each
pair $(U,\epsilon)$ is the required collection.

Now let $A$ be a Lie algebra of global smooth vector fields on $M$
under the usual Lie bracket operation. Let $X_1,...X_k\in A$ and
$\phi^1_{t_1}...\phi^k_{t_k}$ be the local flows associated with
them, for appropriate values of $t_1,...t_k$. Consider the set of
all local diffeomorphisms (where defined) of the form
\begin{equation}
m\rightarrow\phi^1_{t_1}(\phi^2_{t_2}(...\phi^k_{t_k}(m)...))\quad
m\in M
\end{equation}
for each choice of $k\in\N,~X_1,...X_k\in A$ and admissible
$(t_1,...,t_k)\in\R^k$. Now define a relation $\sim$ on $M$ by
$m_1\sim m_2$ if and only if some local diffeomorphism of the form
(2) maps $m_1$ into $m_2$. The relation $\sim$ is, in fact, an
equivalence relation \cite{2,5} and the associated equivalence
classes are called the \emph{orbits} of A. Further, each of these
orbits can be given a unique structure as a connected submanifold
of $M$ \cite{2,5} and is then a \emph{leaf} of $M$ \cite{2}. The
interesting question arises as to whether $A$ is
\emph{integrable}, that is, whether the orbits are integral
manifolds of $A$. More precisely, for each $m\in M$, does
$\Delta(m)$, if non-trivial, coincide with the subspace of $T_mM$
tangent to the orbit of A? Without some further restriction on $A$
the answer is easily shown to be no. Suppose now that $A$ is a
\emph{Lie algebra} of smooth vector fields on $M$ under the Lie
bracket operation. If, in addition, $A$ leads to a Fr\"obenius
type of distribution, so that $\dim\Delta(m)$ is constant on $M$,
the integrability of $A$ follows from Fr\"obenius' theorem (see,
e.g. \cite{1}). However, without the constancy of $\dim\Delta(m)$
the result may fail. To see this the following modified version of
an example in \cite{5} suffices. Let $M=\R^2$ with the usual
global $x,y$ coordinate system and let $f:\R^2\rightarrow\R$ be
the global smooth function given by $f(x,y)=0~(x\leq
0),~f(x,y)=e^{-\frac{1}{x}}~(x>0)$. Then define vector fields
$X_k~(k\in\N)$ on $M$ by
\begin{equation}
X_1=\frac{\partial}{\partial x},~~X_2=f\frac{\partial}{\partial
y},...,X_k=\frac{\partial^{k-2}f}{\partial
x^{k-2}}\frac{\partial}{\partial y},....
\end{equation}
Let $A$ be the vector space spanned by the vector fields in (3).
Then $A$ is a Lie algebra since $[X_1,X_k]=X_{k+1}~(k\geq 2)$ and
$[X_p,X_q]=0~(p,q\geq 2)$. Since the only solution of the equation
$\sum^n_{k=0}a_k\frac{\partial^kf}{\partial x^k}=0$ for
$a_0,...,a_k\in\R$ and for all $x\in\R,~x>0,$ is
$a_0=a_1=...=a_k=0,~A$ is infinite-dimensional. Also,
$\dim\Delta(m)=2$ in the open region $x>0$ of $M$ and
$\dim\Delta(m)=1$ if $x\leq 0$. However, from (2) and (3) it can
be seen that $M$ is the only orbit and has dimension two. Thus $A$
is a Lie algebra but fails to be integrable.

In the above example $A$ was infinite-dimensional. The general
problem of finding integrability conditions for $A$ when
$\dim\Delta(m)$ is not constant over $M$ (i.e. the generalisation
of Fr\"obenius' theorem) has been extensively discussed. The
following result is a consequence of a general result due to
Hermann \cite{6} but draws also on the work of Sussmann \cite{5}
and Stefan \cite{2,7}.

\begin{theorem}
Let $A$ be a finite-dimensional Lie algebra of global smooth
vector fields on a paracompact smooth manifold $M$ with associated
generalised distribution $\Delta$. If $m\in M$ and
$\dim\Delta(m)=0$ the orbit through $m$ is $\{m\}$. If
$\dim\Delta(m)>0$ the orbit through $m$ is a leaf of $M$ and $A$
is integrable in that at each such $m\in M,~\Delta(m)$ coincides
with the subspace of $T_mM$ tangent to the orbit through $m$.
\end{theorem}

\section{Lie algebras of Vector Fields on Space-Times}

Now let $M$ be a space-time. A submanifold $N$ of $M$ of dimension
$1,2$ or $3$ is said to be \emph{timelike} (respectively, \emph{spacelike}
 or \emph{null}) at $m\in N$ if the subspace of $T_mM$
tangent to $N$ is timelike (respectively, spacelike or null). Also
$N$ is called \emph{timelike} (respectively, \emph{spacelike} or
\emph{null}) if it is timelike (respectively, spacelike or null) at
each $m\in N$.

Now let $A$ be a vector space of global smooth vector fields on
$M$ with associated generalised distribution $\Delta$. In the
study of the set $A$ it is convenient to know the variation of the
type and dimension of $\Delta(m)$ over $M$. Thus for $i=0,1,2,3,4$
define the subsets $V_i$ of $M$ by
\begin{equation}
V_i=\{m\in M:\dim\Delta(m)=i\}
\end{equation}
and then for $p=1,2,3$ define the subsets $S_p$, $T_p$ and $N_p$
of $M$ by
\begin{eqnarray}
S_p&=&\{m\in M:\dim\Delta(m)=p\quad\text{and $\Delta(m)$ is spacelike}\}\nonumber\\
T_p&=&\{m\in M:\dim\Delta(m)=p\quad\text{and $\Delta(m)$ is timelike}\}\nonumber\\
N_p&=&\{m\in M:\dim\Delta(m)=p\quad\text{and $\Delta(m)$ is null}\}
\end{eqnarray}
Thus $M=\bigcup^4_{i=0}V_i$ and $V_p=S_p\cup T_p\cup N_p$ for
$p=1,2,3$. The following result gives two ways of disjointly
decomposing $M$ in which, for a subset $U$ of $M$, $\inter U$ means the
topological interior of $U$ with respect to the manifold topology
on $M$. It tidies up the work in \cite{8} where there were careless
slips (see \cite{9} for a corrected version). The following version of
the rank theorem will also be used. If for $m'\in M$
$\dim\Delta(m')=k~(1\leq k\leq 4$ then there exists an open
neighbourhood of $m'$ at each point of which $\dim\Delta(m)\geq
k$.

\begin{theorem}
Let $M$ be a space-time and $A$ a vector space of global smooth
vector fields on $M$. Then, with the above notation, $M$ may be
directly decomposed in the following ways.
\begin{equation}
M=\bigcup^4_{i=0}\inter V_i\cup Z_1=V_4\cup \bigcup^3_{i=0}\inter V_i\cup
Z_1
\end{equation}
\begin{equation}
M=V_4\cup\bigcup^3_{p=1}\inter S_p\cup \bigcup^3_{p=1}\inter T_p\cup
\bigcup^3_{p=1}\inter N_p\cup \inter V_0\cup Z
\end{equation}
where $Z_1$ and $Z$ are closed subsets of $M$ defined by the
disjointness of the decompositions and which have empty interior,
$\inter Z_1=\inter Z=\emptyset$.
\end{theorem}

\begin{proof}
For the first decomposition the rank theorem shows that $V_4$ is
open and hence that $V_4=\inter V_4$. The same theorem also
reveals that the subsets $M_n=\bigcup_{i=n}^4V_i~(0\leq n\leq 4)$
are open in $M$. Now, by the disjointness of (6), $Z_1$ is closed
in $M$ and so let $W\subseteq Z_1$ be an open subset of $M$. By
disjointness, $W\cap V_4=\emptyset$. Then $W\cap M_3$ is open and
equals $W\cap V_3$. If this latter subset is not empty then, since
it is open, it implies that $W\cap \inter V_3\neq \emptyset$. This
contradicts the disjointness of the decomposition (6) and so
$W\cap V_3=\emptyset$. Similarly $W\cap M_2=W\cap V_2$ and is open
and it follows that $W\cap V_2=\emptyset$. Repeating the argument
leads to $W\cap V_i=\emptyset~(0\leq i\leq 4)$ and hence
$W=\emptyset$. Thus $\inter Z_1=\emptyset$.

For the second decomposition it is again a matter of checking that
the (closed) subset $Z$ has empty interior. So let $W\subseteq Z$
be open. As above, $W\cap V_4=\emptyset$ and the subset
$\widetilde{W}$ given by
\begin{equation}
\widetilde{W}\equiv W\cap M_3=W\cap V_3=W\cap(S_3\cup T_3\cup N_3)
\end{equation}
is open. Suppose $\widetilde{W}\neq \emptyset$. If $W$ is disjoint
from $S_3$ \emph{and} $T_3$ one has $W\cap N_3$ open and
non-empty. But this implies that $W\cap \inter N_3\neq\emptyset$
and contradicts the disjointness of the decomposition (7). So $W$
intersects $S_3\cup T_3$ non-trivially. If $W\cap
S_3\neq\emptyset$ let $p\in W\cap S_3\subseteq \widetilde{W}$. Now
$\dim\Delta(m)=3$ at each $m\in \widetilde{W}$ and so there exists
$X,Y,Z\in A$ such that $X(p), Y(p)$ and $Z(p)$ span the spacelike
3-space $\Delta(p)$ at $p$. But this implies that the normal
direction to $\Delta(p)$ is timelike. If $u\in T_pM$ spans this
direction then, in some coordinate neighbourhood $U\subseteq
\widetilde{W}$ of $p$ the components $X^a, Y^a, Z^a$ and $u^a$
satisfy, at $p$, the equations
\begin{equation}
X_au^a=0,~~~Y_au^a=0,~~~Z_au^a=0.
\end{equation}
Suppose, without loss of generality, that the last three columns
of the left hand sides of the array (9) form a non-singular
$3\times 3$ system for $u^1,u^2$ and $u^3$ at $p$. Then, together
with a fixed (constant) choice of $u^0$, Cramer's rule shows that
(9) can be solved for $u^a$ uniquely at $p$ and in some open
neighbourhood $U'\subseteq U$ of $p$ at each point of which $X,Y$
and $Z$ are independent tangent vectors. The components
$u^1,u^2,u^3$ and $u^0$ are then smooth functions on $U'$ and
hence give rise to a timelike vector field (also called u) in some
open neighbourhood $U''(\subseteq\widetilde{W})$ of $p$. It
follows that at each point of $U'', \Delta(m)$ is 3-dimensional
(since $U''\subseteq \widetilde{W}$) and spacelike (since it is
orthogonal to $u$ at each such point). Hence $p\in U''\subseteq
S_3$ and so $U''\subseteq \inter S_3$. This gives the
contradiction $W\cap \inter S_3\neq \emptyset$ to the disjointness
of the decomposition and so $W\cap S_3=\emptyset$. Similarly one
can establish that $W\cap T_3=\emptyset$ and the contradiction
that $W$ is disjoint from $S_3$ and $T_3$. It follows that
$\widetilde{W}=W\cap M_3=\emptyset$. One then repeats the argument
on the open subsets $W\cap M_2$ and then $W\cap M_1$ to finally
achieve $W=\emptyset$ and the conclusion that $\inter
Z=\emptyset$.
\end{proof}

\section{Killing Symmetry in Space-Times}

Let $M$ be a space-time and let $K(M)$ be the vector space of all
global smooth \emph{Killing vector fields} on $M$ and with
associated generalised distribution $\Delta$. Then if $X\in K(M)$
one has $\Lie_X g=0$ and thus in any coordinate domain in $M$
\begin{equation}
X_{a;b}+X_{b;a}=0~~~(\Leftrightarrow X_{a;b}=F_{ab}=-F_{ba})
\end{equation}
\begin{equation}
X^a{}_{;bc}=F^a{}_{b;c}=R^a{}_{bcd}X^d
\end{equation}
where $F$ is the \emph{Killing bivector} of $X$. It easily follows
that (10) and (11) give rise along any curve in $M$ to a system of
first order differential equations in the ten quantities $X^a$ and
$F_{ab}$. Since $M$ is path connected each $X\in K(M)$ is uniquely
determined by the values of $X$ and $F$ at some (any) point of $M$
(and so $\dim K(M)\leq 10$). Hence if $X$ vanishes on some
non-empty open subset of $M,~X\equiv 0$ on $M$ (and thus $\inter
V_0=\emptyset$ in (6) and (7) if $A\equiv K(M)$ is non-trivial).
It is also well known that $X,Y\in K(M)\Rightarrow [X,Y]\in K(M)$
and so $K(M)$ is a finite-dimensional Lie algebra called the {\sl
Killing algebra} of $M$. It follows from theorem 1 that $K(M)$ is
integrable so that, in the sense of that theorem, the orbit
structure is well behaved.

Now suppose $M$ is decomposed as in (4)-(7) with $K(M)$ playing
the role of $A$ in these equations. First consider the subset
$V_4$ of $M$ and suppose $V_4\neq \emptyset$. Then $V_4$ is an
open submanifold of $M$ with Lorentz metric induced from the
metric $g$ on $M$. Let $K(V_4)$ be the restriction to $V_4$ of the
members of the Killing algebra $K(M)$. The generalised
distribution on $V_4$ associated with $K(V_4)$ has dimension four
at each point and so from Chow's theorem \cite{10} (see also
\cite{2,5}) the associated orbits of $K(V_4)$ are precisely the
components of $V_4$. Hence the orbits of $K(M)$ through any point
of $V_4$ are the components of $V_4$.

Now let $O$ be a \emph{proper} orbit of $K(M)~(i.e.~1\leq \dim
O\leq 3)$. Then, from theorem 1, $O$ is a leaf of $M$. Now let $f$
be any local diffeomorphism of the type (2) arising from $K(M)$
and with domain the open subset $U$ of $M$ and range the open
subset $U'\equiv f(U)$ of $M$. Suppose $U\cap O\neq\emptyset$.
Then $f:U\rightarrow M$ is smooth and since $O\cap U$ is an (open)
submanifold of $O$ and hence of $U$, $f$ restricts to a smooth map
$O\cap U\rightarrow M$ whose range lies in the leaf $O$ of $M$.
Thus (section 2) it restricts to a smooth map $O\cap U\rightarrow
O$ and hence to a smooth map $f:O\cap U\rightarrow O\cap U'$ since
$O\cap U'$ is an open and hence regular manifold of $O$. Thus, in
this sense, each such map $f$ restricts to a local diffeomorphism
of $O$. It now follows (section 2) that if $m\in O\cap U$ and
$m'\equiv f(m)\in O\cap U'$ and if $u,v\in T_pM$ are each tangent
to $O$ at $m$ then $f_*u$ and $f_*v$ are tangent to $O$ at $m'$.
Since the statement that $X$ is a Killing vector field on $M$ is
equivalent to each local flow $\phi_t$ of $X$ satisfying the local
isometry condition $\phi^*_tg=g$, where $\phi^*_t$ is the usual
\emph{pullback} map, one has
\begin{equation}
g_{m'}(f_*u,f_*v)=(f^{-1*}g)_{m'}(f_*u,f_*v)=g_m(u,v)
\end{equation}
where $g_m=g(m)$. From this it easily follows from the definitions
of section 1 that \emph{the nature (spacelike, timelike or null) of
$O$ is the same at each point of $O$}.

Now suppose that $O$ is proper and non-null and let
$i:O\rightarrow M$ be the inclusion map. Since $O$ is connected
(theorem 1) it easily follows from the previous paragraph that $O$
is either spacelike or timelike. Then $g$ induces a metric
$h\equiv i^*g$ on $O$ which is a positive definite metric if $O$
is spacelike and a Lorentz metric if $O$ is timelike. Since $K(M)$
is integrable, any $X\in K(M)$ is tangent to $O$ and projects
naturally to a smooth global vector field $\widetilde{X}$ on $O$
such that $i_*\widetilde{X}=X$ on $O$ \cite{1}. Also if $\phi_t$ is a
local flow (isometry) of $X$ and $\widetilde{\phi}_t$ the
corresponding local flow on $O$ associated with $\widetilde{X}$
then $i\circ\widetilde{\phi}_t=\phi_t\circ i$. To see this note
that if $\widetilde{c}$ is an integral curve of $\widetilde{X}$
then $c\equiv i\circ\widetilde{c}$ is an integral curve of $X$
\cite{1}. Conversely, if $c$ is an integral curve of $X$ and if
$\widetilde{c}$ is defined by $c=i\circ\widetilde{c}$ then
$\widetilde{c}$ is an integral curve of $\widetilde{X}$. This
follows because $c:I\rightarrow M$ is smooth for some open
interval $I$ of $\R$ and its range lies in the leaf $O$ and is
represented by the map $\widetilde{c}$, which is thus smooth. The
condition that $c$ is an integral curve of $X$ is $X\circ
c=c_*\circ\frac{\partial}{\partial t}$ (with $t$ the parameter of
$c$) and so $c_*\circ\frac{\partial}{\partial
t}=(\underline{i}\circ\widetilde{c})_*\circ\frac{\partial}{\partial
t}=i_*\circ\widetilde{c}_*\circ\frac{\partial}{\partial t}=X\circ
c=X\circ
i\circ\widetilde{c}=i_*\circ\widetilde{X}\circ\widetilde{c}$.
Since $i$ is an immersion (and so $i_*$ is injective),
$\widetilde{c}_*\circ\frac{\partial}{\partial
t}=\widetilde{X}\circ\widetilde{c}$ and so $\tilde{c}$ is an
integral curve of $\widetilde{X}$. Now since $\phi_t^*g=g$ one has
\begin{equation}
\widetilde{\phi}_t^*h=\widetilde{\phi}^*_t(i^*g)
=(i\circ\widetilde{\phi}_t)^*g=(\phi_t\circ i)^*g=i^*(\phi^*_tg)=h
\end{equation}
and so $\widetilde{X}$ is a \emph{Killing vector field on $O$ with
respect to $h$}. Also, if $X,Y\in K(M)$ with associated vector
fields $\widetilde{X}$ and $\widetilde{Y}$ on $O$ a standard
result on Lie brackets gives
$i_*[\widetilde{X},\widetilde{Y}]=[X,Y]$. It follows that the map
$k:X\rightarrow \widetilde{X}$ is a Lie algebra homomorphism
$K(M)\rightarrow K(O)$ where $K(O)$ is the Lie algebra of global
smooth Killing vector fields on $O$.

The linear map $k$ need not be either injective or surjective. To
see that it is not necessarily surjective consider the space-time
$M=\{(t,x,y,z)\in \R^4,~t>0\}$ with metric given by
\begin{equation}
ds^2=-dt^2+tdx^2+e^{2t}dy^2+e^{3t}dz^2
\end{equation}
Here $K(M)$ is 3-dimensional being spanned by
$\frac{\partial}{\partial x},\frac{\partial}{\partial y}$ and
$\frac{\partial}{\partial z}$. Any orbit $O$ of $K(M)$, however,
is a 3-dimensional spacelike hypersurface of constant $t$ which,
with its induced metric, is 3-dimensional Euclidean space. Hence
$\dim K(O)=6$.

To see that $k$ need not be injective one first notes that lack of
injectivity would require $X\in K(M),~X\not\equiv 0$, with
$\widetilde X\equiv 0$ on $O$. This means that $X$ should vanish
on $O$. To investigate this possibility let $X\in
K(M),~X\not\equiv 0$ be such that $X(m)=0$ for some $m\in M$. The
associated local flows $\phi_t$ associated with $X$ then satisfy
$\phi_t(m)=m$ and $m$ is called a \emph{zero} (or a \emph{fixed
point}) of $X$. Let $U$ be a coordinate neighbourhood of $m$ with
coordinates $y^a$. The linear isomorphisms
$\phi_{t*}:T_mM\rightarrow T_mM$ are represented in the basis
$(\frac{\partial}{\partial y^a})_m$ by the matrix (or its
transpose depending on notation)
\begin{equation}
e^{tB}=\exp{t}(B^a_b),~~B^a_b=(\frac{\partial X^a}{\partial
y^b})_m
\end{equation}
where $B$ is the \emph{linearisation} of $X$ at $m$. Since $X\in
K(M)$ it follows from (10) that $B^a_b=F^a{}_b(m)$. Now $X$ is an
\emph{affine} vector field and so if $\chi$ is the usual \emph{exponential}
diffeomorphism from some open neighbouhood of $o\in
T_mM$ onto some open neighbourhood $V$ of $m$ then \cite{3}
\begin{equation}
\phi_t\circ\chi=\chi\circ\phi_{t*}
\end{equation}
In the resulting normal coordinate system on $V$ with coordinates
$x^a$ it follows that one has the convenient situation that the
components $X^a$ of $X$ are \emph{linear} functions of the
$x^a,~X^a=B^a_bx^b$ (see, e.g. \cite{11,12}). Since
$B^a_b(=F^a{}_b(m))$ is \emph{skew self-adjoint} with respect to
$g(m)$ (i.e. $B_{ab}\equiv g_{ac}B^c_b=-B_{ba}$ at $m$) \emph{the
rank of $B$ is even}. If $B=0$ then a remark at the beginning of
this section shows that $X\equiv 0$ on $M$. Thus the rank of $B$
is two or four. The zeros of $X$ in $V$ have coordinates
satisfying $B^a_bx^b=0$ and so, if $\rank B=4$, the zero $m$ of
$X$ is the only zero in $V$ (and so is \emph{isolated}). If,
however, $\rank B=2$ \emph{the set of zeros of $X$ in $V$ can be
given the structure of a 2-dimensional regular manifold $N$ of the
open submanifold $V$ of $M$} \cite{11,12}. Thus if $k$ is not
injective with $X$ vanishing on the orbit $O$ through $m$, the
zeros of $X$ are not isolated since, in $V$, they are precisely
the points of $N$. The next step is to get some information on the
dimension of $O$ using the discussion in section 2. To do this let
$O'=O\cap V$ so that $O'$ is an open subset, and hence an open
submanifold, of $O$ and hence a submanifold of $M$. Then $O'$ is a
submanifold of $M$ contained in the regular submanifold $V$ of $M$
and is then a submanifold of $V$. By the definition of $N$ it then
follows that $O'\subseteq N\subseteq V$ with $O'$ and $N$
submanifolds of $V$ with $N$ regular. From this it follows that
$O'$ is a submanifold of $N$ and so $\dim O'\leq \dim N$. This
finally gives $\dim O(=\dim O')\leq 2$. Thus if $\dim O=3,~k$ is
injective.

The above results can now be collected together.

\begin{theorem}
Let $M$ be a space-time and let $O$ be an orbit of $K(M)$.
\begin{enumerate}
\item If $\dim O=4,~O$ is a component of $V_4$ in the decomposition
(6) and the map $k:K(M)\rightarrow K(O)$ is an injective Lie
algebra homomorphism.

\item If $1\leq \dim O\leq 3$ then O is spacelike, timelike or
null. If, in addition, $O$ is not null, the map $K:K(M)\rightarrow
K(O)$ is a Lie algebra homomorphism which is not necessarily
surjective. It is injective if $\dim O=3$ but need not be if $\dim
O=1$ or $2$. If $k$ is injective then $K(M)$ is Lie-isomorphic to
a Lie subalgebra of $K(O)$.
\end{enumerate}
\end{theorem}

\begin{proof}
The proof has mostly been given above. To complete the proof of
(i) one notes that injectivity follows since if $X\in K(M)$
vanishes on the open subset $O$ then $X\equiv 0$. In (ii) the last
sentence is clear and the proof is completed by the following
examples. First let $M_1=\R^2$ with positive definite metric $g_1$
given by $e^{x^2+y^2}(dx^2+dy^2)$ so that $K(M_1)$ is
1-dimensional and spanned by the Killing vector field
$x\frac{\partial}{\partial y}-y\frac{\partial}{\partial x}$ which
has a single zero at the origin. Next let $M_2=(-1,\infty)\times
\R$ with Lorentz metric $g_2$ given in coordinates $t,z$ by
$-dt^2+2dtdz+tdz^2$ so that $K(M_2)$ is 1-dimensional and spanned
by the nowhere-zero Killing vector field $\frac{\partial}{\partial
z}$. Note that, in this case, the Killing orbits are 1-dimensional
and are timelike for $-1<t<0$, null for $t=0$ and spacelike for
$t>0$. Finally let $M_3=\R^2$ with the usual 2-dimensional
Minkowski metric $g_3$ so that $K(M_3)$ is 3-dimensional. Now let
$(M',g')$ be the \emph{metric product} of $(M_1,g_1)$ and
$(M_2,g_2)$ so that $M'=M_1\times M_2$ and $g'=g_1\otimes g_2$.
Then $K(M')$ can be checked to be the vector space sum
$K(M_1)\oplus K(M_2)$ and is thus 2-dimensional. In $M'$, the
submanifolds of the form $N'=\{(0,0)\}\times N$, where $N$ is any
Killing orbit in $M_2$, are each 1-dimensional Killing orbits in
$M'$ which are spacelike (respectively, timelike or null) if the
orbit $N$ is spacelike (respectively, timelike or null). From the
choice of $(M_2,g_2)$, each can occur. All other orbits in $M'$
are 2-dimensional and may be spacelike, timelike or null. Again,
each possibility can occur. Restricting to the situation when the
1-dimensional orbits $N'$ are non-null one sees that the
associated map $k:K(M')\rightarrow K(N')$ is not injective. Next
let $(\widetilde{M},\widetilde{g})$ be the metric product of
$(M_1,g_1)$ and $(M_3,g_3)$. Again $K(\widetilde{M})=K(M_1)\oplus
K(M_3)$ and so $\dim K(\widetilde{M})=4$ and
$\widetilde{N}\equiv\{(0,0)\}\times M_3$ is a 2-dimensional
timelike orbit with all other orbits timelike and 3-dimensional.
Clearly $\widetilde{N}$ is flat and $\dim K(\widetilde{N})=3$ and
so the map $k:K(\widetilde{M})\rightarrow K(\widetilde{N})$ is not
injective. An obvious variant of this argument using the metric
product of Euclidean space $\R^2$ with the manifold $\R^2$ with
Lorentz metric $e^{-t^2+z^2}(-dt^2+dz^2)$ and with single
independent Killing vector field $z\frac{\partial}{\partial
t}+t\frac{\partial}{\partial z}$ yields an example of a space-time
$\overline{M}$ with $\dim K(\overline{M})=4$ and which admits a
single 2-dimensional flat spacelike orbit and with all other
orbits 3-dimensional and either timelike $(z^2>t^2)$, spacelike
$(z^2<t^2)$ or null $(t=\pm z)$.
\end{proof}

The reason that $k$ may fail to be injective is because \emph{a
set of Killing vector fields on $M$ may be independent as members
of the vector space $K(M)$ but restrict to a dependent set of
members of $K(O)$ for some orbit $O$}. This is what happens in the
examples above where non-trivial members of $K(M)$ actually vanish
on $O$. This will be explored further in theorem 7.

Now define a subset $I_mK(M)$ (or, simply, $I_m$ if $K(M)$ is
clear) of $K(M)$ for $m\in M$ by
\begin{equation}
I_mK(M)~(=I_m)=\{X\in K(M):X(m)=0\}
\end{equation}
It is clear that $I_m$ is a vector subspace of $K(M)$ and, in
fact, a \emph{subalgebra} (the isotropy subalgebra) of $K(M)$
since $X(m)=0$ and $Y(m)=0\Rightarrow[X,Y](m)=0$. Also, if $X\in
K(M)$ and $X(m)=0$, then, by the above definition of $K(M)$, each
associated map $\phi_t$ of $X$ satisfies $\phi_t(m)=m$ and also
the derived linear map $\phi_{t*}:T_mM\rightarrow T_mM$ is a
member of the Lorentz group with respect to $g(m)$. If one chooses
a coordinate neighbourhood $U$ of $p$ with coordinates $x^a$, and
hence a basis $(\frac{\partial}{\partial x^a})_m$ for $T_mM$, a
basis for the Lie algebra of this Lorentz group in this basis is
the 6-dimensional Lie algebra (under matrix commutation) of
matrices $A^a_b$ satisfying $g_{ac}A^c_b+g_{bc}A^c_a=0$ where
$g_{ab}$ are the components of $g(m)$ in these coordinates. Thus
the map $f$ which associates $X\in I_m$ with its Killing bivector
$F^a{}_b(m)(=X^a_{,b}(m)$-see (15)) can be regarded as a (linear)
map from $I_m$ to the Lie algebra of the Lorentz group. Further,
if $Y$ is also in $I_m$ with Killing bivector $G$, $f$ maps
$Y\rightarrow G^a{}_b(m)$ and then $f$ maps $[X,Y]$ to
$(F^a{}_cG^c{}_b-G^a{}_cF^c{}_b)(m)$. Thus $f$ is a Lie algebra
homomorphism from $I_m$ to the Lie algebra of the Lorentz group
and, since $X(m)=0,~F^a{}_b(m)=0\Rightarrow X\equiv 0$ on $M,~f$
is an isomorphism onto its range. Thus \emph{$I_m$ is isomorphic
to a subalgebra of the Lie algebra of the Lorentz group}. Also,
for any $m\in M$, the map $g:K(M)\rightarrow T_mM$ given by
$X\rightarrow X(m)$ is linear with range $\Delta(m)$ (see(1)) and
kernel $I_m$. Since $\dim\Delta(m)=dimO$, where $O$ is the orbit
through $m$, it follows that $\dim K(M)=\dim O+\dim I_m$.

If $m\in M$ is a zero of $X\in K(M)$ then a certain amount of
information is available about the Ricci and Weyl tensors at $m$.
This information depends on the subgroup $I_m$ of the Lorentz
group, that is, on the Killing bivectors of those members of
$K(M)$ which vanish at m. The following theorem summarises the
situation, the proof of which can be found in \cite{12}-\cite{14}.
It is remarked here that if $0\not\equiv X\in K(M)$ with
$X(m)=0~(m\in M)$ then the relationship $\pounds_XR_{ab}=0$
evaluated at $m$ using (10) gives $F_a{}^cR_{cb}+F_b{}^cR_{ca}=0$.
Thus the 2-space of $T_mM$ representing the blade of $F$ (if $F$
is simple at $m$) or the pair of 2-spaces of $T_mM$ representing
the canonical pair of blades of $F$ (if $F$ is not simple at $m$)
are each eigenspaces of the Ricci tensor and hence the latter is,
in this sense, degenerate \cite{15}.

\begin{theorem}
Let $M$ be a space-time, let $m\in M$ and suppose that $I_m$ is
not trivial. Then the Petrov type at $m$ is $O$, $N$ or $D$ and
the Ricci tensor admits at least one eigenvalue degeneracy. In
detail, and identifying members of $I_m$ with their bivectors at
$m$,
\begin{enumerate}
\item if any of these (simple or non-simple) bivectors is non-null
the Petrov type at $m$ is $O$ or $D$ whilst if any is null the
Petrov type at $m$ is $O$ or $N$ (and in either case the Segre
type of the Ricci tensor is degenerate), \item if $\dim I_m=3$ the
Weyl tensor vanishes at $m$ and the Ricci tensor has Segre type
either $\{1,(111)\},\{(1,11)1\}$ or $\{(211)\}$ (or the degeneracy
of each of the first two of these types), \item if $\dim I_m\geq
4$ then at $m$ the Weyl tensor is zero and the Ricci tensor
satisfies $R_{ab}=R/4g_{ab}$.
\end{enumerate}
\end{theorem}

\section{The Orbit Structure of $K(M)$}

Here the relationship between $K(M)$ and the possible dimensions
of the associated orbits will be investigated. To do this a
distinction must be made between those orbits which are `stable'
with respect to their nature or dimension (or both) and those
which are not. Let $O$ be a proper orbit. Then either $O\subseteq
S_p,~O\subseteq T_p$ or $O\subseteq N_p$ for some $p,~1\leq p\leq
3$. Call $O$ \emph{stable} if $O$ is entirely contained in either
$\inter S_p$, $\inter T_p$ or $\inter N_p$ or, equivalently, if
for each $m\in O$ there exists a neighbourhood $U$ of $m$ such
that all orbits through all points of $U$ are of the same nature
and dimension as $O$. Thus if $O$ is stable then nearby orbits
have the same nature (spacelike, timelike or null) and dimension
as $O$. Next call $O$ \emph{dimensionally stable} if $O$ is
entirely contained in some $\inter V_p~(1\leq p\leq 3)$ or,
equivalently, if for $m\in O$ there is an open neighbourhood $U$
of $O$ such that all orbits through all points of $U$ are of the
same dimension as $O$. Clearly if $O$ is stable it is
dimensionally stable. Thus, for example, in the extended
Schwarzschild solution one has 3-dimensional Killing orbits
everywhere which are timelike and stable outside the Schwarzschild
radius, spacelike and stable inside this radius and null and
unstable (but dimensionally stable) on it. For the generic F.R.W.
metrics the orbits are everywhere 3-dimensional spacelike and
stable whilst for the usual Einstein static universe there is a
single 4-dimensional orbit. Of the examples described in the proof
of theorem 3, the space-time $M'$ has the submanifolds $N'$ as
1-dimensional dimensionally unstable (and hence unstable) orbits
whilst all others are 2-dimensional and each is stable except the
null one (which is, however, dimensionally stable). The space-time
$\widetilde {M}$ has $\widetilde {N}$ as an unstable (and
dimensionally unstable) orbit with all other stable and the
space-time $\overline{M}$ admits a 2-dimensional spacelike
unstable (and dimensionally unstable) orbit with all others
3-dimensional and each of these stable except the null one (which
is, however, dimensionally stable).

\begin{theorem}
Let $M$ be a space-time.
\begin{enumerate}
\item If $O$ is an orbit of $K(M)$ such that $O\cap \inter
S_p\neq\emptyset$ (respectively, $O\cap \inter T_p\neq\emptyset$
or $O\cap \inter N_p\neq\emptyset$) for some $p~(1\leq p\leq 3)$,
then $O\subseteq \inter S_p$ (respectively, $O\subseteq \inter
T_p,~O\subseteq \inter N_p$) and is hence stable. Thus the open
dense subset $M\setminus Z$ of $M$ in the decomposition (7) is the
union of all stable orbits of $M$ whilst the nowhere dense closed
subset $Z$ is the union of all the unstable orbits of $M$.
Similarly if $O\cap \inter V_p\neq\emptyset~(1\leq p\leq 3)$ then
$O\subseteq \inter V_p$ and $O$ is dimensionally stable. Thus the
open dense subset $M\setminus Z_1$ of $M$ in the decomposition (6)
is the union of all dimensionally stable orbits of $M$ whilst the
nowhere dense closed subset $Z_1$ is the union of all
dimensionally unstable orbits. \item In the decomposition (7) of
$M,~S_3$ and $T_3$ are open subsets of $M$. Thus each
3-dimensional spacelike or timelike orbit of $K(M)$ is stable.
\end{enumerate}
\end{theorem}

\begin{proof}
(i) Let $O$ be such an orbit with $O\cap \inter S_p\neq\emptyset$
and let $m\in O\cap \inter S_p$. If $m'$ is any other point of $O$
there exists an open neighbourhood $U$ of $m$ and a local isometry
$f$ of the form (2) which is defined on $U$ such that $~f(m)=m'$
and $U\subseteq \inter S_p$. The orbits intersecting $U$ are
$p-$dimensional and spacelike and similarly for each point in the
open neighbourhood $f(U)$ of $m'$. Thus $O\subseteq \inter S_p$
and a similar argument suffices for the timelike and null cases.
The rest of the proof is similar.
\par \noindent
\par \noindent
(ii) Let $O$ be a 3-dimensional spacelike orbit. If $m\in O$
construct a Gauss coordinate domain $U$ about $m$ such that each
point of $U$ lies on a time-like geodesic which intersects $O$
orthogonally. Denoting the tangent vector of such a geodesic by
$K(t)$, with $t$ an affine parameter, one has the well-known
result (see e.g. \cite{16}) that if $X\in K(M)$ then $X^ak_a$ is
constant along the geodesic (and hence zero along the geodesic
since it vanishes on $O$). Thus at any point of $U$ the orbit of
$K(M)$ is orthogonal to a timelike vector and is hence spacelike
with dimension at most three. However, the rank theorem ensures
that $U$ can be reduced (if necessary) so that the orbit through
any point of $U$ is of dimension at least three. Thus each orbit
through $U$ is 3-dimensional and spacelike and so $U\subseteq
\inter S_3$ and $O$ is stable. A similar argument applies in the
timelike case and so $S_3=\inter S_3$ and $T_3=\inter T_3$ in (7).
\end{proof}

It is remarked that, as a consequence of theorem 5(i), there is an
open neighbourhood of any point in the open dense subset
$M\setminus Z$ (i.e. of "almost any" point of $M$) in which the
orbits are of the same dimension and nature.

It was mentioned at the beginning of section 4 that $\dim K(M)\leq
10$. If $\dim K(M)$
\par \noindent
\par \noindent
$=10$ then $M$ is of constant curvature and if $M$ is of constant
curvature then each $m\in M$ admits a connected neighbourhood $U$
such that, with its induced metric, $U$ satisfies $\dim K(U)=10$.
If $\dim K(M)=9$ then, as will be seen later (theorem 6), the only
possible orbits are 4-dimensional and so $M=V_4$ from theorem 2.
But then a result at the end of section 4 shows that $\dim
I_m=9-4=5$ for each $m\in M$ and this is a contradiction since the
Lorentz group has no 5-dimensional subgroups. Thus $\dim K(M)=9$
is impossible. If $\dim K(M)=8$ then, again, theorem 6 will show
that $M=V_4$ and hence that $\dim I_m=4$ at each $m\in M$. It
follows from theorem 4 that $M$ is of constant curvature and so,
locally, admits a 10-dimensional Killing algebra. Thus, if $m\in
M$, there exists an open neighbourhood $U$ of $m$ such that $\dim
K(U)=10$. (c.f Yegerov's theorem quoted in \cite{17}). It should
be remarked at this point that the impossibility of $\dim K(M)=9$
may be thought an immediate consequence of Fubini's theorem
\cite{18} and which appears to rule out $\dim K(M)=1/2n(n+1)-1$
for a manifold $M$ of dimension $n$ and with metric $g$. As far as
the present author can tell, Fubini's theorem only deals with the
positive definite case but, fortunately, can be extended to cover
(almost) all signatures and, in particular, applies to space-times
(see appendix).

\begin{theorem}
Let $M$ be a space-time. Then the following hold for the orbits of
$K(M)$.
\begin{enumerate}
\item If there exists a 3-dimensional null orbit, then $3\leq \dim
K(M)\leq 7$. If, however, there exists a null dimensionally stable
3-dimensional orbit or any non-null 3-dimensional orbit then
$3\leq\dim K(M)\leq 6$.
\item If there exists a 2-dimensional null orbit, then $2\leq\dim
K(M)\leq 5$ and if there exists a 2-dimensional non-null orbit,
$2\leq\dim K(M)\leq 4$. If there exists any 2-dimensional
dimensionally stable orbit, $2\leq\dim K(M)\leq 3$.
\item If there exists a 1-dimensional null orbit, $1\leq \dim
K(M)\leq 5$ and if there exists a 1-dimensional non-null orbit,
$1\leq\dim K(M)\leq 4$. If there exists a 1-dimensional
dimensionally stable orbit, $\dim K(M)=1$.
\end{enumerate}
\end{theorem}

\begin{proof}
(i) If any 3-dimensional orbit $O$ exists then $\dim K(M)\geq 3$.
Suppose, in addition, that $O$ is null (and not necessarily
dimensionally stable). Let $m\in O$ and choose independent members
$X',Y'$ and $Z'$ in $K(M)$ which span $O$ at $m$. Then in some
coordinate neighbourhood $U$ of $m$ the smooth vector field $k$ in
$U$ with components $k^a=\varepsilon^a{}_{bcd}X'^bY'^cZ'^d$, where
$\varepsilon$ denotes the usual alternating symbol, is orthogonal
to $X',Y'$ and $Z'$ on $U$. Also, since $X',Y'$ and $Z'$ span $O$
at $m,~U$ may (and will) be chosen so that none of $X',Y',Z'$ and
$k$, can vanish at any point of $U$. Thus, since $O$ is null, $k$
is null and tangent to $O$ on $O'\equiv U\cap O$. Now let $X$ be a
non-trivial member of $K(M)$. Then $X$ is tangent to $O$ and hence
orthogonal to $k$ on $O'$. Thus $(X^ak_a)_{;b}p^b=0$ on $O'$ where
$p^a$ are the components of any tangent vector to any curve in $M$
lying in $O'$. Suppose $X(m)=0$. Then evaluating this differential
condition at $m$, using (10), shows that the Killing bivector $F$
of $X$ satisfies $F_{ab}p^ak^b=0$ at $m$ for any $p\in T_mM$
tangent to $O$. It follows from elementary bivector algebra that
$F_{ab}k^b=\lambda k_a$ at $m~(\lambda\in \R)$ and that at most
four independent bivectors at $m$ can have this property (since
$k$ is null). Thus $\dim I_m\leq 4$ and so $\dim K(M)\leq 7$.

Now suppose the orbit $O$ is 3-dimensional and either null
\emph{and} dimensionally stable or non-null (and hence necessarily
stable by theorem 5). Let $m\in O$ and $U$ an open neighbourhood
of $m$ such that each orbit intersecting $U$ has the same
dimension and type as $O$. By the same argument as in the previous
paragraph, one can (reducing $U$ if necessary) choose a smooth
vector field $k$ on $U$ everywhere orthogonal to the orbits in
$U$. Thus if $X$ is a non-trivial member of $K(M),~X^ak_a=0$ on
$U$. Now suppose $X(m)=0$ and evaluate the condition
$(X^ak_a)_{;b}=0$ at $m$ to see that the Killing bivector $F$ of
$X$ satisfies $F_{ab}k^b=0$ at $m$. Since only three independent
bivectors $F$ may have this property one sees that $\dim I_m\leq
3$ and hence that $\dim K(M)\leq 6$. [In fact if $O$ is
\emph{non-null},
 the map $K(M)\rightarrow K(O)$ given by
$X\rightarrow\widetilde{X}$ is injective (theorem 3) and the above
result follows in this case ( since $\dim K(O)\leq 6)$ without
appeal to the dimensional stability of $O$.]
\par \noindent
\par \noindent
(ii) If there exists a 2-dimensional orbit $O$ then $\dim K(M)\geq
2$. Let $m\in O$ and let $Y,Z\in K(M)$ such that $Y$ and $Z$ span
$O$ at $m$. Now let $U$ be a coordinate neighbourhood of $m$ such
that $G_{ab}\equiv 2Y_{[a}Z_{b]}$ is a smooth simple nowhere-zero
bivector on $U$ and (hence so is its dual $G^*_{ab}$). It follows
(see, e.g. \cite{19}) that, reducing $U$ if necessary, one may write
$G^*_{ab}=2P_{[a}Q_{b]}$ for some smooth vector fields $P$ and $Q$
on $U$ and then $P$ and $Q$ span the orthogonal complement of the
tangent space to $O$ at each point of $O\cap U$. Now suppose $\dim
K(M)>2(=\dim O)$. Then, as in the proof of part (i), there exists
$X\in K(M),~X\not\equiv 0$, such that $X(m)=0$. Also, since $X$ is
tangent to $O,~X^aP_a=X^aQ_a=0$ on $O$. Thus
\begin{equation}
(X^aP_a)_{;b}Y^b=(X^aP_a)_{;b}Z^b=(X^aQ_a)_{;b}Y^b=(X^aQ_a)_{;b}Z^b=0
\end{equation}
On evaluating (18) at $m$ one finds that the bivector $F$ of $X$
satisfies, at $m$, the conditions
\begin{equation}
F_{ab}P^aY^b=F_{ab}P^aZ^b=F_{ab}Q^aY^b=F_{ab}Q^aZ^b=0
\end{equation}
If $O$ is non-null, (19) represents four independent conditions on
$F$ at $m$ and thus there are only two independent bivectors in
the vector space of bivectors at $m$ satisfying them. In this case
(c.f part (i)) one finds $2\leq \dim K(M)\leq 4$. If $O$ is null,
(19) represent only three restrictions (since in this case the
tangent space to $O$ is a null 2-space and intersects its
orthogonal complement in a null direction and so $X,Y,P$ and $Q$
do not yield independent tangent vectors at any point of $O$).
Thus in this case there are exactly three independent bivectors
satisfying (19) and so $2\leq \dim K(M)\leq 5$. If $O$ is a
dimensionally stable orbit, then by following closely the proof of
part (i) one has for $m\in O,~X\in K(M),~X\not\equiv 0,~X(m)=0$,
the consequences $X^aP_a=X^aQ_a=0$ on some open neighbourhood of
$m$ and so
\begin{equation}
(X^aP_a)_{;b}=(X^aQ_a)_{;b}=0~~(\Rightarrow
F_{ab}P^b=F_{ab}Q^b=0~at~m)
\end{equation}
There is only one independent bivector satisfying the bracketed
consequence in (20) and so $\dim I_m\leq 1$. Hence $2\leq \dim
K(M)\leq 3$.
\par \noindent
\par \noindent
(iii) The proof of (iii) is similar. If $O$ is a 1-dimensional
orbit and if $m\in O$ choose $X\in K(M)$ which spans $O$ at $m$
and a coordinate neighbourhood $U$ of $m$ in which
$X=\partial/\partial x^1$. Then the three independent vector
fields on $U$ whose associated covector fields on $U$ are
$dx^a~(a=2,3,4)$ are orthogonal to $X$ on $U$. The result now
follows with the last part of it being simply the statement that
if $X,Y\in K(M)$ and $X=\phi Y$ over some non-empty open subset
$U$ of $M$ for some function $\phi:U\rightarrow \R$, then $\phi$
is constant and $X=\phi Y$ on $M$ \cite{16}.
\end{proof}

Regarding theorem 6(i) the author does not know of an example of a
space-time $M$ with $\dim K(M)=7$ and where a dimensionally
unstable 3-dimensional null orbit (or orbits) exist. If such an
$M$ exists then theorem 6 shows that all other orbits are
4-dimensional and clearly $V_0=\emptyset$ in the decomposition (7)
otherwise for $m\in V_0$ one would obtain the contradiction $\dim
I_m=7$. Thus, in (7), one would have $M=V_4\cup Z$ where $Z=N_3$
(and $\inter Z=\emptyset$).

Also regarding theorem 6(i) the situation with $\dim K(M)=5$ and
$O$ a 3-dimensional spacelike orbit is impossible since, for $m\in
O$, one would have $I_m$ isomorphic to a 2-dimensional subalgebra
of $so(3)$, which does not exist. The non-homogeneous plane waves
show, however, that $\dim K(M)=5~(or~6)$ and $O$ a 3-dimensional
null (and stable) orbit is possible. Now consider the situation
when $\dim K(M)=5$ and $O$ is a 3-dimensional timelike
(necessarily stable) orbit. It is sometimes argued that this
implies that $\dim K(O)=5$ and that this is impossible by
appealing to Fubini's theorem. However, theorem 3 shows that $\dim
O\geq 5$ and it can then be established that $O$ is a
3-dimensional manifold of constant curvature and so $O$ admits a
\emph{local} 6-dimensional Killing algebra about each of its
points. It is not immediately obvious that this contradicts the
fact that $\dim K(M)=5$. Also, the above method of contradicting
the situation when $O$ is spacelike fails here since $so(1,2)$
admits a 2-dimensional subalgebra. To proceed further consider the
open submanifold $T_3$ of $M$ where the orbits are timelike and
3-dimensional and let $m\in T_3$. Then $I_m$ is a 2-dimensional
subalgebra of $o(1,2)$ generating the subgroup of null rotations
of the 3-dimensional Lorentz group. This is easily checked either
by direct computation or by noting that it must generate a
subgroup of the 4-dimensional Lorentz group which fixes a
spacelike vector (the normal to the orbit) and then checking the
possibilities from this latter Lorentz group (see, e.g
\cite{20,21}). Next, using an argument similar to that in the
proof of theorem 6, one can construct a coordinate neighbourhood
$U\subseteq T_3$ of $m$ and a smooth geodesic vector field $Y$ on
$U$ which is everywhere orthogonal to the orbits and which gives
Gauss coordinates on $U$ in which the metric is
\begin{equation}
ds^2=dy^2+g_{\alpha\beta}dx^\alpha dx^\beta
\end{equation}
$(\alpha,\beta=0,1,2)$ and where
$Y^a=\delta^a_3,~Y_a=y_{,a}=\delta_a^3,~Y^aY_a=1$ and
$Y_{a;b}Y^b=0$. At each point $m'$ of $U$ one may extend $Y(m')$
to a null tetrad $l,n,x,Y(m')$ and then $I_m$ may be regarded as
being spanned by the bivectors $l_{[a}n_{b]}$ and $l_{[a}x_{b]}$.
Theorem 4 and the remarks preceding it can then be used to show
that $T_3$ is a conformally flat region and that, in $U$, the
Ricci tensor takes the form
$R_{ab}=\alpha(y)g_{ab}+\beta(y)Y_aY_b$ (where, since each subset
of $U$ of constant $y$ lies in an orbit, the Ricci eigenvalues
$\alpha$ and $\beta$ depend, at most, on $y$). The conformally
flat Bianchi identity is
\begin{equation}
R_{c[a;b]}=\frac{1}{6}g_{c[a}R_{,b]}
\end{equation}
where $R\equiv R_{ab}g^{ab}$ is the Ricci scalar and a contraction
of (22) with $Y^a$ yields
\begin{equation}
6\beta
Y_{a;b}=(2\dot{\alpha}-\dot{\beta})(g_{ab}-Y_aY_b)~~(.\equiv d/dy)
\end{equation}
In the open subregion $A\subseteq T_3$ on which $\beta\neq 0$ one
may solve (23) for $Y_{a;b}$ and by a standard argument [use (21)
and (23) to get
$Y_{a;b}=-\Gamma^3_{ab}\Rightarrow(\frac{2\dot{\alpha}
-\dot{\beta}}{6\beta})g_{\alpha\beta}=1/2g_{\alpha\beta,3}\Rightarrow
g_{\alpha\beta}=h(y)q_{\alpha\beta}$ for a positive function
$h(y)$ and functions $q_{\alpha\beta}$ independent of $y$] cast
the metric locally into the form
\begin{equation}
ds^2=dy^2+S^2(y)q_{\alpha\beta}dx^\alpha dx^\beta
\end{equation}
for a smooth function $S$ and where $q_{\alpha\beta}$ are the
components of a (3-dimensional constant curvature) Lorentz metric
in the orbit through $m$. The metric (24) is just the spacelike
equivalent of a F.R.W. metric and it easily follows that $T_3$
admits, locally, a Lie algebra of Killing vector fields of
dimension at least six. If $B$ is the subregion of $T_3$ on which
$\beta=0$ then (23) shows that $\alpha$ is constant on each
component of $\inter B$ and so $\inter B$ is an open subregion of
$M$ on which the Einstein space and conformally flat conditions
hold and so is of constant curvature (and so the local Killing
algebra is 10-dimensional). Thus, for $\dim K(M)=5$, each point of
the open dense subset $A\cup \inter B$ of $T_3$ admits, locally, a
Killing algebra of dimension greater than five. Thus from the
physical viewpoint, if one wishes to study symmetry corresponding
to $\dim K(M)$ being \emph{exactly} five (even locally) then on
$T_3$ (and on $M$ if all orbits are 3-dimensional and timelike)
such conditions cannot be achieved.

All the other dimensionally stable possibilities in theorem 6 can
be constructed \cite{14}. The examples given after theorem 3 confirm
the existence of space-times with $\dim K(M)=4$ and $\dim O=2$
with $O$ not dimensionally stable and either timelike or
spacelike, and with $\dim K(M)=2$ and $\dim O=1$ with $O$ not
dimensionally stable and either timelike, spacelike or null. Now
consider the space-time
metrics
\begin{equation}
ds^2=-dt^2+e^{x^2+y^2+z^2}(dx^2+dy^2+dz^2)
\end{equation}
\begin{equation}
ds^2=dz^2+e^{-t^2+x^2+y^2}(-dt^2+dx^2+dy^2)
\end{equation}
each on the manifold $R^4$. For each of these space-times $\dim
K(M)=4$ with $K(M)$ being spanned by the global Killing vector
fields with components (in the coordinates $t,x,y,z$) given by
$(1,0,0,0),(0,0,z,-y),(0,y,-x,0)$ and $(0,z,0,-x)$ for (25) and by
$(0,0,0,1),(x,t,0,0),(y,0,t,0)$ and $(0,-y,x,0)$ for (26). Then
(25) admits the submanifold $x=y=z=0$ as a 1-dimensional timelike
orbit which is not dimensionally stable, with all other orbits
stable, 3-dimensional and timelike. Similarly, (26) admits the
submanifold $t=x=y=0$ as a 1-dimensional spacelike orbit which is
not dimensionally stable, with all other orbits stable,
3-dimensional and timelike. Thus many of the dimensionally
unstable possibilities in theorem 6 parts (ii) and (iii) can
occur. There are, however, some that cannot. For example, if $\dim
K(M)=3$ and $O$ is a 1-dimensional timelike orbit then, for $m\in
O$, one again has the contradiction that $I_m$ is a 2-dimensional
subalgebra of $so(3)$. The other situations allowed by this
theorem are, at least to the author, as yet unresolved.

\section{Further Remarks}

For space-times admitting Killing symmetry and for which a
non-trivial isotropy $I_m$ exits, much valuable information is
available from the bivectors at $m$ associated with the members of
$I_m$. At this point it is useful to note (by adopting the
discussion of section 4) that on a 2-dimensional manifold a
non-zero bivector is necessarily a non-singular matrix and so a
zero of a non-trivial Killing vector field on such a manifold is
necessarily isolated. Similarly, on a 3-dimensional manifold, a
non-zero bivector has rank equal to two and is hence a singular
matrix. Thus a zero of a non-trivial Killing vector field in this
case is never isolated but, rather (should a zero exist), its
zeros constitute (locally) a 1-dimensional submanifold.

From the calculations in the proof of theorem 6, and given $m\in
M$ and the nature (spacelike, timelike or null) and dimension of
the orbit $O$ through $p$ (assumed proper) and whether $O$ is
dimensionally stable or not, it is straightforward to write down a
basis for the vector space of Killing bivectors at $p$ for members
of $I_m$. (It is not claimed that all possibilities not so far
eliminated from theorem 6 can exist.) From this information (or by
an easy direct calculation) it turns out that whatever the nature
of $O$, provided it is dimensionally stable, any Killing bivector
at $m$ arising from a non-trivial member $X$ of $I_m$ is
\emph{simple} and its blade is tangent to $O$ at $m$ (and note
that if $I_m$ is not trivial then $O$ cannot be 1-dimensional).
Thus $m$ is not isolated.

For an orbit $O$ which is \emph{not} dimensionally stable the
calculations for theorem 6 again show that the Killing bivector at
$m\in O$ arising from some $X\in I_m$ is also simple (and so $m$
is not isolated) except, possibly, in the cases when $O$ is either
1-dimensional and null, 2-dimensional and non-null or
3-dimensional and null. For example, consider the space-time
$\widetilde{M}=M_1\times M_3$ constructed during the proof of
theorem 3 and let $X$ be non-trivial member of $K(M_1)$ which
vanishes at the origin $o$ of $M_1$. Now let $Y$ be the vector
field on $M_3$ given in the obvious coordinates by
$Y=z\partial/\partial t+t\partial/\partial z$. Then $Y$ vanishes
at the origin $o'$ of $M_3$ and the vector field sum $X\oplus Y$
is a member of $K(\widetilde{M}$ which vanishes at
$(o,o')\in\widetilde{M}$ and its Killing bivector there is easily
checked to be non-simple. In this case $\dim K(\widetilde{M})=4$
and the orbit through $(o,o')$ is 2-dimensional, timelike and not
dimensionally stable. Finally, consider the example of the
space-time $M=R^4$ with metric
\begin{equation}
ds^2=e^{-t^2+x^2}(-dt^2+dx^2)+e^{y^2+z^2}(dy^2+dz^2)
\end{equation}
Here a basis for $K(M)$ can be taken as consisting of the vector
fields $X=t\partial/\partial x+x\partial/\partial t$ and
$Y=y\partial/\partial z-z\partial/\partial y$ and so $\dim
K(M)=2$. The Killing vector field $X+Y$vanishes at the origin and
its bivector there is non-simple. The origin is itself a
(non-proper) orbit and the isotropy subalgebra there is
2-dimensional. This example also possesses 1-dimensional
spacelike, timelike and null orbits which are not dimensionally
stable in the submanifolds $\{(t,x,0,0)~t,x\in R,~t^2+x^2\neq 0\}$
and similar spacelike ones in the submanifolds $\{(0,0,y,z)~y,z\in
R,~y^2+z^2\neq 0\}$. All other orbits are 2-dimensional, stable
and can be spacelike, timelike or null.

It was mentioned, following the proof of theorem 3, that an
independent set of Killing vector fields on $M$ when restricted to
an orbit $O$ of $K(M)$ may no longer be independent. However, in
all the examples given of this phenomenon, $O$ was not
dimensionally stable or 3-dimensional. This feature turns out to
be general as the next theorem shows.

\begin{theorem}
Let $M$ be a space-time and let $X^1,...,X^n$ be independent
members of $K(M)$. If an orbit $O$ is 3-dimensional or is any
dimensionally stable orbit of $K(M)$ then the restrictions
$\widetilde{X^1},...,\widetilde{X^n}$ to $O$ are independent
vector fields on $O$ (and hence independent members of $K(O)$ if
$O$ is non-null). Said another way, a non-trivial Killing vector
field cannot vanish on such an orbit $O$ (that is, the Lie algebra
homomorphism $X\rightarrow \widetilde{X}$ is injective).
\end{theorem}

\begin{proof}
Let $X$ be a non-trivial member of $K(M)$ and let $O$ be an orbit
of $K(M)$ such that $X$ vanishes on $O$. It follows immediately
from theorem 3(ii) that $O$ cannot be 3-dimensional and from
theorem 6(iii) that $O$ cannot be dimensionally stable and
1-dimensional. So let $O$ be dimensionally stable with $\dim O=2$
and, following the proof of theorem 6 since $O$ is dimensionally
stable, for $m\in O$ let $U$ be a coordinate neighbourhood of $m$
in $M$ such that there exist $Y,Z\in K(M)$ with $Y(m')$ and
$Z(m')$ spanning the tangent space to $O$ at each $m'\in U$ and
such that there exist smooth vector fields $W$ and $T$ on $U$
which are orthogonal to the (2-dimensional) orbits of $K(M)$ at
each point of $U$. Then $X^aW_a=X^aT_a=0$ on $U$ and so
differentiating these equations and evaluating on $O$ (where $X$
vanishes) gives $F_{ab}W^b=F_{ab}T^b=0$ at each point of $O$,
where $F$ is the Killing bivector of $X$. But since $X$ vanishes
on $O$ it also follows that, on $O,~X^a{}_{;b}Y^b=X^a{}_{;b}Z^b=0$
and so, on $O$, $F_{ab}Y^b=F_{ab}Z^b=0$. It is now clear that,
whether $O$ is null or non-null, $F$ (and $X$) vanishes on $O$ and
hence one has the contradiction that $X\equiv 0$ on $M$. The
result follows.
\end{proof}

\renewcommand{\theequation}{A\arabic{equation}}
\setcounter{equation}{0}
\section*{Appendix (Fubini's Theorem)}

For a manifold $M$ of dimension $n$ and with metric $g$ of
arbitrary signature a theorem, often quoted in the literature and
due to Fubini \cite{18},  says that the Lie algebra $K(M)$ cannot
have dimension $\frac{1}{2}n(n+1)-1$ (that is, one less than the
maximum for that dimension). However, the proof in \cite{18} (see
also \cite{16}) and its reliance on Bianchi's (positive definite)
results recently reprinted in \cite{22} suggest that Fubini's
theorem should be revisited. It was, of course, shown earlier that
Fubini's theorem holds for a space-time $M$, that is, $K(M)$
cannot be $9$-dimensional. To see that Fubini's theorem holds for
almost all dimensions and signatures consider the orthogonal (Lie)
group $SO(p,q)~(p+q=n)$ where, because of the symmetry isomorphism
$SO(p,q)\approx SO(q,p)$, one need only consider the case $p\leq
q$. This group has dimension $\frac{1}{2}n(n-1)$ and acts on $R^n$
as a Lie transformation group in the usual way \cite{1}. It can be
realised as the symmetry group of the metric $\gamma$ with
components $\diag(-1,...,-1,1,...1)$ where there are $p$ minus and
$q$ plus signs. If $H$ is a Lie subgroup of $SO(p,q)$ then, by
restriction, $H$ also acts on $R^n$ as a Lie transformation group.
Thus if $0\neq v\in R^n$ and if $H_v$ is the Lie subgroup of $H$
which fixes $v$ and $O_v$ is the orbit of $v$ under $H$, one has
\cite{1}
\begin{equation}
\dim H=\dim H_v+\dim O_v
\end{equation}
Now for $n=3$ and for $n\geq 5$ it is known \cite{23}  that if $H$
is a \emph{proper} Lie subgroup of $SO(n),~\dim
H\leq\frac{1}{2}(n-1)(n-2)$. It is not true for $n=4$ since
$SO(4)$ has a 4 (but no higher)-dimensional subgroup. The
following lemma is now useful (c.f. \cite{24}).

\begin{lemma}
Let $H$ be a Lie subgroup of $SO(p,q)$ with $p\leq q,~p+q=n\geq
3$. If for each spacelike $v\in R^n,~H_v\approx SO(p,q-1)$ or for
each timelike $v\in R^n,~H_v\approx SO(p-1,q)$, then $H\approx
SO(p,q)$.
\end{lemma}

\begin{proof}
Let $x_1,...,x_n$ be an appropriate orthonormal basis for $R^n$.
If $p=0$ (the positive definite case, so that each $x_k$ is
spacelike) then each $H_{x_k}\approx SO(n-1)$. Thus the skew
self-adjoint (with respect to $\gamma$) matrices represented in an
obvious notation by $x_i\wedge x_j~(i<j,i\neq k\neq j)$ are in the
Lie algebra of $H_{x_k}$ since if $G$ is any of these matrices,
$Gx_k=0$. Since this is true for each $k$, the Lie algebra of $H$
coincides with that of $SO(n)$ and the result follows. Now suppose
that $p\geq 1$ so that one may assume $x_1$ timelike and $x_2$ and
$x_3$ spacelike. If $H_v\approx SO(p-1,q)$ for each timelike $v$
then, with $v=x_1$, the matrices $x_i\wedge x_j$ for $2\leq i< j$
are in the Lie algebra of $H_v$ and hence of $H$. Now
$x'=x_1+1/2x_2$ is timelike and for $i\geq 3$ the matrices
$G_i=(x_1+2x_2)\wedge x_i$ satisfy $G_ix'=0$ and so are in the Lie
algebra of $H$ for each $i\geq 3$. Hence so are the matrices
$x_1\wedge x_i~(i\geq 3)$. A similar consideration of the timelike
vector $x_1+1/2x_3$ then shows that $x_1\wedge x_2$ is in the Lie
algebra of $H$ and hence that $H\approx SO(p,q)$. A similar
argument reveals the same conclusion if $H_v\approx SO(p,q-1)$ for
each spacelike $v$.
\end{proof}

\begin{lemma}
If $H$ is a proper Lie subgroup of any of the Lie groups
$SO(p,q),~p\leq q,~p+q=n,~q\geq 3$, then $\dim
H<1/2n(n-1)-1$.
\end{lemma}

\begin{proof}
The remarks following (A1) show the result to be true for
$SO(n)~(n\neq 4)$. That it holds for $SO(4)$ follows by choosing
$0\neq v\in R^4$ such that $H_v$ is a \emph{proper} subgroup of
$SO(3)$ (from lemma 1) and so $\dim H_v\leq 1$. Since $\dim
O_v=3$, (A1) shows that $\dim H\leq 4$. Now consider $SO(1,n-1)$
for $n\geq 4$ and choose a timelike $v$ in $R^n$ for which $H_v$
is a \emph{proper} subgroup of $SO(n-1)$. Equation (A1) and the
previous result for $SO(n)$ show that
\begin{equation}
\dim H<1/2(n-1)(n-2)-1+n-1=1/2n(n-1)-1
\end{equation}
An induction argument on $p$ completes the
proof.
\end{proof}

Thus lemma 2 holds for all signatures except $(0,2),(1,1),(1,2)$
and $(2,2)$ and its failure in the first two of these is clear. It
also fails for the signature $(1,2)$ because of the existence of a
2-dimensional Lie subgroup of the 3-dimensional Lorentz group. To
see why it fails for signature $(2,2)$ let $t_1,t_2,x_1$ and $x_2$
be an orthonormal basis for $R^4$ with $t_1,t_2$ timelike and
$x_1,x_2$ spacelike. Then change to a new basis $l,n,x_1x_2$ with
$l\equiv t_1+x_1$ and $n\equiv t_2+x_2$ orthogonal null vectors.
Next consider the subspace $W$ of the Lie algebra $o(2,2)$ of
$SO(2,2)$ spanned by $l\wedge n,~l\wedge x_1,~l\wedge x_2,~n\wedge
x_1$ and $n\wedge x_2$. It is easily checked that $W$ is a
5-dimensional subalgebra by either directly computing the
appropriate commutators or, more quickly, by noting that if
$A,B\in W$ then $[A,B]l$ and $[A,B]n$ are linear combinations of
$l$ and $n$ and hence $[A,B]$, when decomposed in the obvious
basis has no term in $x_1\wedge x_2$. The unique connected
5-dimensional Lie subgroup of $SO(2,2)$ associated with $W$ is
precisely the subgroup which fixes the \emph{totally null}
2-dimensional subspace $N$ of $R^4$ spanned by $l$ and $n$ (that
is each non-zero member of $N$ is null).

The following `Fubini theorem' can now be
proved.

\begin{theorem}
Let $M$ be a connected smooth paracompact manifold of dimension
$n\geq 3$ admitting a smooth metric $g$ of signature $(p,q)$ with
$p\leq q$ and $q\geq 3$. Suppose $\dim K(M)\neq 1/2n(n+1)$. Then
$\dim K(M)<1/2n(n+1)-1$.
\end{theorem}

\begin{proof}
Let $m\in M$ and $O$ be the orbit associated with $K(M)$ through
$m$. Then $O$ is a smooth integral manifold of $K(M)$ and $\dim
K(M)=\dim O+\dim K^*_m$. If $\dim O<n$, arguments such as those
used in the proof of theorem 6 show that $K^*_m$ is restricted and
so $\dim K^*_m<1/2n(n-1)-1$ from lemma 2. It follows that if $\dim
K(M)$ is not the maximum possible then
\begin{equation}
\dim K(M)<n+1/2n(n-1)-1=1/2n(n+1)-1
\end{equation}
\end{proof}

The signatures not covered by the theorem are, again
$(0,2),(1,1),(1,2)$ and $(2,2)$. In the first two of these the
final conclusion of the theorem is false since the ordinary
cylinder and the `Lorentz cylinder' obtained, respectively, by
identifying the points $(x,t)$ and $(x+k,t)$ for each $k\in \Z$,
in $\R^2$, on which the standard Euclidean and Minkowski metrics
have been placed, have a 2-dimensional Killing algebra. However,
if a 2-dimensional manifold $M$ has $\dim K(M)=2$, it is easily
checked from theorem 6 that the orbits are 2-dimensional over an
open dense subset $A$ of $M$ and then \cite{16}  that $A$ and
hence $M$ have constant curvature. Thus the Killing algebra is
\emph{locally} 3-dimensional. In the $(1,2)$ and $(2,2)$ cases the
situation is, to the best of the author's knowledge, unresolved.
In the $(1,2)$ case, if $\dim K(M)=5$, one again has constant
curvature on $M$ and hence a \emph{local} 6-dimensional Killing
algebra \cite{25}. It is easily checked from \cite{26,27}  that in
the $(1,1),(0,2)$ and $(1,2)$ signatures, \emph{if $M$ is simply
connected}, $\dim K(M)\neq 1/2n(n+1)-1$.

It is remarked that the result in lemma 2 can be
strengthened.

\begin{lemma}
Let $H$ be a proper Lie subgroup of $SO(p,q),~p\leq
q,~p+q=n,~q\geq 5$. Then $\dim H\leq 1/2(n-1)(n-2)+p$. The result
also holds for the signatures $(0,2),(0,3),(1,2),(1,3)$ and
$(2,2)$.
\end{lemma}

The proof uses the result for $SO(n)$ quoted earlier and then
essentially the argument given in lemma 2 and proceeds by
induction from equations like (A1) so that for $v$ timelike
$(p\geq 1)$
\begin{equation}
\dim H\leq n-1+1/2(n-2)(n-3)+p-1=1/2(n-1)(n-2)+p
\end{equation}
This proof and that of lemma 2 are modelled on the work in
\cite{24} where lemma 3 for Lorentz signature was established. A
further strengthening of lemma 2 and the Fubini theorem can then
be achieved.

\section*{Acknowledgments}
The author acknowledges valuable discussions with Dimitri
Alexeevski, Bartolom$\acute{e}$ Coll, Michael Crabb and Alan
Rendall and thanks M. Sharif for help in preparing the manuscript.


\newpage





\begin{thebibliography}{99}

\bibitem{1} F. Brickell and R.S. Clark. {\sl Differential
Manifolds}, Van Nostrand, (1970).

\bibitem{2} P. Stefan. Proc. London Math. Soc. {\bf 29}, (1974),
699.

\bibitem{3} S. Kobayashi and K. Nomizu. {\sl Foundations of
Differential Geometry} Vol 1, Interscience, New York, (1963).

\bibitem{4} R. Abraham, J.E. Marsden and T. Ratiu. {\sl Manifolds,
Tensor Analysis and Applications}. Springer 2nd edition, (1988)

\bibitem{5} H.J.Sussmann. Trans. Am. Math. Soc. {\bf 180}, (1973),
171.

\bibitem{6} R. Hermann. International Symposium on Nonlinear
Differential Equations and Nonlinear Mechanics, Academic Press,
New York, (1963), 325.

\bibitem{7} P. Stefan. J. London Math. Soc. {\bf 21}, (1980), 544.

\bibitem{8} G.S. Hall. Class. Quant. Grav. {\bf 13}, (1996), 1479.

\bibitem{9} G.S. Hall. The Petrov Lectures in the Proceedings of
the Volga Summer School XII, University of Kazan, (2000).

\bibitem{10} W.L. Chow. Math. Ann. {\bf 117}, (1939), 98.

\bibitem{11} G.S. Hall. Gen. Rel. Grav. {\bf 20}, (1988), 671.

\bibitem{12} G.S. Hall. J. Math. Phys. {\bf 31}, (1990), 1198.

\bibitem{13} J. Ehlers and W. Kundt. in {\sl Gravitation: An
Introduction to Current Research} ed. L. Witten. Wiley, New York,
(1962).

\bibitem{14} H. Stephani, D. Kramer, M.A.H. MacCallum, C. Hoenselaers
and E. Herlt. {\it Exact Solutions of Einstein's Field Equations},
Cambridge University Press, (2003).

\bibitem{15} G.S. Hall and C.B.G. McIntosh. Int. J. Theor. Phys.
{\bf 22}, (1983), 469.

\bibitem{16} L.P. Eisenhart. {\sl Riemannian Geometry}, Princeton
University Press, (1966).

\bibitem{17} A.Z. Petrov. {\sl Einstein Spaces}, Pergamon, (1969).

\bibitem{18} G.Fubini. Annali. di. Matematica. {\bf 8}, (1903), 39.

\bibitem{19} G.S. Hall and A.D. Rendall. Int. J. Theor. Phys. {\bf
28}, (1989), 365.

\bibitem{20} R. Shaw. Quart. J. Math. Oxford {\bf 21}, (1970), 101.

\bibitem{21} G.S. Hall {\sl Curvature Structure and Symmetries in
General Relativity}, World Scientific. To appear.

\bibitem{22} L. Bianchi. Gen. Rel. Grav. {\bf 33}, (2001), 2171.

\bibitem{23} D. Montgomery and H. Samelson. Anns. Maths. {\bf 44},
(1943), 454.

\bibitem{24} A.D. Rendall. Class. Quan. Grav. {\bf 5}, (1988), 695.

\bibitem{25} G.S. Hall and M.S. Capocci. J. Math. Phys. {\bf 40},
(1999), 1466.

\bibitem{26} K. Nomizu. Anns Maths. {\bf 72},
(1960), 105.

\bibitem{27} G.S. Hall.  Class. Quant. Grav. {\bf 6},
(1989), 157.

\end{thebibliography}
\end{document}